\documentclass[letter]{aa}  
\usepackage{natbib}

\usepackage{graphicx}
\usepackage{txfonts}
\begin{document}

   \title{First confirmed ultra-compact dwarf galaxy \\in the NGC\,5044 group}

   \author{Favio R. Faifer
          \inst{1,2},
         Carlos G. Escudero\inst{1,2},
         Mar\'ia C. Scalia\inst{1,2},
         Anal\'ia V. Smith Castelli\inst{2,3},
         Mark Norris\inst{4},
         Mar\'ia E. de Rossi\inst{5,6},
         Juan C. Forte\inst{3,7},
         and Sergio A. Cellone\inst{1,2}
         }

   \institute{Facultad de Ciencias Astron\'omicas y Geof\'isicas,
      Universidad Nacional de La Plata, Paseo del Bosque s/n, B1900FWA, 
      La Plata, Argentina\\
             \email{favio@fcaglp.unlp.edu.ar}
         \and
             Instituto de Astrof\'isica de La Plata 
             (CCT La Plata - CONICET - UNLP), Paseo del Bosque s/n, 
             B1900FWA, La Plata, Argentina
        \and Consejo Nacional de Investigaciones Cient\'ificas y T\'ecnicas, 
              Godoy Cruz 2290, C1425FQB, CABA, Argentina
        \and Jeremiah Horrocks Institute, University of Central Lancashire, Preston, PR1 2HE, United Kingdom
        \and Universidad de Buenos Aires, Facultad de Ciencias Exactas y Naturales y Ciclo Básico Común. Buenos Aires, Argentina
        \and CONICET-Universidad de Buenos Aires, Instituto de Astronomía y Física del Espacio (IAFE).
        \and Planetario `Galileo Galilei', Secretar\'ia de Cultura, 
        Av. Sarmiento S/N, 1425 Ciudad Autónoma de Buenos Aires, Argentina
        }

   \date{Received March 1, 2017; accepted March 16, 2017}

 
  \abstract
   {Ultra-compact dwarfs (UCDs) are stellar systems displaying
colours and metallicities between those of globular clusters (GCs) and 
early-type dwarf galaxies, as well as sizes of $R_{\rm eff} \lesssim 100$ pc and 
luminosities in the range $-13.5<M_{\rm V}<-11$ mag. Although their origin is 
still subject of debate, the most popular scenarios suggest that they are 
massive star clusters or the nuclei of tidally stripped dwarf galaxies.}
   {NGC\,5044 is the central massive elliptical galaxy of the NGC\,5044
group. Its GC/UCD system is completely unexplored. }
   {In Gemini+GMOS deep images of several fields around NGC\,5044 
and in spectroscopic multi-object data of one of these fields, 
we detected an unresolved source with $g'\sim20.6$ mag, compatible 
with being an UCD. Its radial velocity was obtained with FXCOR and
the penalized pixel-fitting (pPXF) code. To study its stellar 
population content, we measured the Lick/IDS 
indices and compared them with predictions of single stellar population models, 
and we used the full spectral fitting technique. }
   {The spectroscopic analysis of the UCD revealed a radial velocity that 
agrees with the velocity of the elliptical galaxy NGC\,5044. From 
the Lick/IDS indices,  we have obtained a luminosity-weighted age and 
metallicity of $11.7^{+1.4}_{-1.2}$ Gyr and $[Z/H]=-0.79\pm0.04$ dex, 
respectively, as well as $[\alpha/Fe]=0.30\pm0.06$. From the full spectral
fitting technique, we measured a lower age (8.52 Gyr) and a similar
total metallicity ($[Z/H]= -0.86$ dex).}  
   {Our results indicate that NGC\,5044-UCD1 is most likely an extreme GC ($M_V\sim
-12.5$ mag) belonging to the GC system of the elliptical galaxy NGC\,5044. }


\authorrunning{Faifer et al.}
\titlerunning {First confirmed UCD in NGC\,5044}
   \maketitle
%

\section{Introduction}
\label{introduccion}

Ultra-compact dwarfs (UCDs) are stellar systems displaying masses,
colours, and metallicities within the range covered by 
those of globular clusters (GCs) and dwarf galaxies. They show typical 
effective radii ($R_{\rm eff}$) $\lesssim $100 pc and luminosities in the 
range $-13.5<M_{\rm V}<-11$ mag \citep{2006RMxAC..26Q.194M}. 
It now seems likely that they are a mixed population of massive star clusters 
and
the nuclei of tidally stripped dwarf galaxies (i.e. \citealp{2009RvMA...21..199H,2011AJ....142..199B,2011MNRAS.414..739N,2014MNRAS.444.3670P,2014prpl.conf..291L,2015ApJ...802...30Z}). However, it has so far proved difficult to conclusively demonstrate
that any individual UCD is of either type (and in particular,
that it is of massive star cluster type).

Initially, UCDs were identified in the environments of rich 
galaxy clusters like Virgo and Fornax. However, more recently, several authors 
have reported UCDs in groups of galaxies and associated with galaxies in the field 
\citep{2007MNRAS.378.1036E,2009MNRAS.394L..97H,2011MNRAS.414..739N,2014MNRAS.443.1151N}. The existence of such objects
in low-density environments provides an important insight into the relative frequency
of the different formation pathways. However, deep spectroscopy, which gives  
information about the stellar population content among other properties, is still scarce \citep{2012MNRAS.425..325F,2016MNRAS.456..617J}. 

We present the spectroscopic confirmation of the first UCD
in the vicinity of the massive elliptical galaxy NGC\,5044 
($M_{\rm B}=-21.2$ mag, $D=35.7\pm5$  Mpc, $V_{\rm r}=2782$ km s$^{-1}$), the 
central object of the NGC\,5044 group. This group harbours $\sim150$ members, 
most of which are dwarf galaxies \citep{1990AJ....100....1F}. This population represents 
half the members of the Fornax Cluster. However, its central
density and velocity dispersion are higher, while its kinematics and early-type
galaxy fraction reveal a mature group \citep{2005MNRAS.356...41C}. Although the elliptical galaxy NGC\,5044 has been the subject of several studies, its rich 
GC system ($\sim5500 \pm 500$ members, Faifer et al. in preparation) remains 
completely unexplored. 

Through the analysis of
Gemini+GMOS images and spectra, we have been able to obtain the photometric 
parameters and the radial velocity of the UCD, as well as information about 
its stellar population content. The letter is organized as follows. In
Section\,\ref{Observations} we present our data, in Section\,\ref{Results}
our analysis and results, and in Section\,\ref{Conclusions} we discuss the results and we present the conclusions.   


\section{Observational data}
\label{Observations}

With the Gemini Multi-Object Spectrograph (GMOS)
\citep{2004PASP..116..425H} of GEMINI-South, we obtained deep
$g'$, $r'$, and $i'$ images of several fields around NGC\,5044 
(Program: GS-2009A-Q-46). In this letter we focus on the field that
contains the elliptical galaxy. The analysis of the complete photometric dataset 
will be presented in a forthcoming paper (Faifer et al. in preparation). 
GMOS consists  of three CCDs of 2048 $\times$ 4096 pixels, 
separated by gaps of $\sim$ 2.8 arcsec, 
with an unbinned pixel scale of 0.0727 arcsec pixel$^{-1}$.  The
field of view (FOV) is 5.5' $\times$ 5.5', and the scale for binning 2$\times$2 
is $\sim$ 0.146 arcsec pixel$^{-1}$.
The images were reduced in the usual way \citep{2015MNRAS.449..612E}.

\begin{table}
\caption{Log of the GEMINI-GMOS data.}
\label{info}
\centering
\begin{tabular}{cccc@{}}
\hline\hline
\multicolumn{1}{c}{Date} & \multicolumn{1}{c}{Filter/Grism} & \multicolumn{1}{c}{Exposures} & \multicolumn{1}{c}{FWHM/Resolution}\\
\multicolumn{1}{c}{} & \multicolumn{1}{c}{} & \multicolumn{1}{c}{(sec)} & \multicolumn{1}{c}{(arcsec/\AA)}\\
\hline
27 Jan. 2009 & g' & 11$\times$450  &  0.92 \\
             & r' &  2$\times$350  &  0.76 \\
             & i' &  7$\times$350  &  0.80 \\
26 Feb. 2009 & g' &  4$\times$450  &   \\
             & r' &  2$\times$350  &   \\
22 Mar. 2010 & B600\_G5323  & 3$\times$2330 & 4.61 \\ 
19 Apr. 2010 & B600\_G5323  & 3$\times$2330 & 4.61 \\ 
20 Apr. 2010 & B600\_G5323  & 2$\times$2330 & 4.61 \\ 
\hline
\end{tabular}
\end{table}

We have also obtained a GMOS spectroscopic mask of the field containing
NGC\,5044 with the aim at obtaining radial velocities of a sample of GC
candidates (program: GS-2010A-Q-56). The instrumental setup included  
2$\times$2 binning and slits of 1$''$ width, with dithering in the spectral 
direction in order to fill the CCD gaps. The data were 
reduced using the IRAF tasks \textsf{gbias},  \textsf{gsflat}, 
\textsf{gsreduce,} and \textsf{mgswavelength}. The individual spectra were 
then extracted using \textsf{apall} and were combinded through the task 
\textsf{scombine}. Finally, an approximate flux calibration was applied based 
on the flux of the standard star LTT\,4816. The final spectrum covers a spectral 
range of 3750-6780 $\AA$ and displays a resolution of $\sim 4.61$ $\AA$. 
 The
signal-to-noise ratio (S/N) per $\AA$ of the final spectrum increases from 20 at 4100 $\AA$ to 70 at 6000 $\AA$. In Table\,\ref{info} we show the basic information related with all our data.

\section{Analysis and results}
\label{Results}

\begin{table}
\caption{Basic information of NGC\,5044-UCD1. 
Apparent magnitudes and colours are not extinction or reddening corrected.}
\label{parameters}
\centering
\begin{tabular}{@{}lc@{}r@{}}
\hline\hline
R.A. (J2000) & 13:15:35.75  &        \\
DEC (J2000)  & -16:23:25.33 &          \\
$A_g$ (mag) & 0.231 & \\                
$A_r$ (mag) & 0.160 & \\
$A_i$ (mag) & 0.119 & \\
$m_{\rm g'}$ (mag) & 20.62  &  $\pm$ 0.01  \\
$(g'-r')$ (mag) & 0.67  &   $\pm$ 0.01  \\
$(g'-i')$ (mag) & 0.98  &   $\pm$ 0.01  \\
$(r'-i')$ (mag) & 0.31  &   $\pm$ 0.01  \\
$V{\rm r}$ (km s$^{-1}$) & 2945   $\pm$ 5.00 \\
$Age_{Lick}$ (Gyr) & 11.7 & $\pm$  1.4\\
$[Z/H]_{Lick}$ & -0.79 & $\pm$  0.04\\
$[\alpha/Fe]$ & 0.30 & $\pm$ 0.06\\
$M_{ssp}$ $(M_\odot)$ & $2.8\times10^7 $& $\pm 3\times10^6$ \\ 
$Age_{pPXF}$ (Gyr) & 8.52 & $\pm$  ---\\
$[Z/H]_{pPXF}$ & -0.86 & $\pm$  ---\\
$H_{\delta_A}$ (\AA) & 1.46 & $\pm$  0.36\\
$H_{\delta_F}$ (\AA) & 1.53 & $\pm$  0.24\\
$G_{4300}$ (\AA) & 3.59 & $\pm$ 0.28\\
$Fe_{4383}$ (\AA) &   2.47 & $\pm$ 0.36\\
$H_\beta$ (\AA)  &  2.18 & $\pm$ 0.12\\
$Fe_{5015}$ (\AA) &  3.01 & $\pm$ 0.25\\
$Mgb$ (\AA)  &  2.35 & $\pm$ 0.11\\
$Fe_{5270}$ (\AA) & 1.68 & $\pm$ 0.13\\
$Fe_{5335}$ (\AA) & 1.69 & $\pm$ 0.14\\
$Fe_{5406}$ (\AA) & 0.74 & $\pm$ 0.11\\
$[MgFe]'$ (\AA) &  1.99 & $\pm$ 0.11 \\
$\langle Fe \rangle$ (\AA) & 1.69 & $\pm$ 0.10\\
\hline
\end{tabular}
\end{table}

\subsection{Photometry}
\label{photometry}

We have subtracted the galaxy light from our images using an iterative 
combination of the background-modelling tool of SExtractor 
\citep{2010ascl.soft10064B} and a median filtering, following 
\citet{2004A&A...415..123P}.
SExtractor was also used to perform the detection of GC candidates in the
field. At the distance adopted here for NGC\,5044, 1 arcsec is equivalent to 
170 pc. This means that any globular cluster or UCD present in the field will 
appear as an unresolved source in our images. We have used \textsf{Daophot} 
to obtain the photometry of the sources detected in the field, and we modelled
the psf using $\sim$ 25-30 well-exposed unresolved objects.

\begin{figure}
\centering
\includegraphics[scale=0.33]{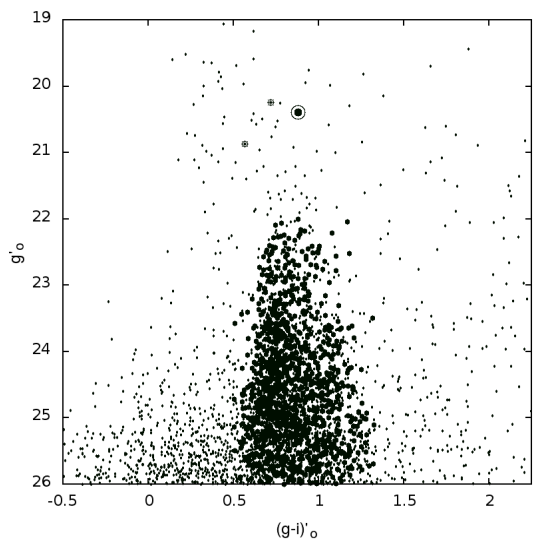}
\caption{Colour-magnitude diagram of the unresolved sources detected in 
the field of NGC\,5044 (dots). Small circles depict the objects
that match our colour selection for globular clusters. The small circles with a 
cross inside show two candidates whose radial velocities proved that they are 
MW stars. The circle surrounded by the ring shows the location 
of NGC\,5044-UCD1 in the diagram.}
\label{color-magnitude}
\label{cmd}
\end{figure}

Figure \ref{cmd} shows the colour-magnitude diagram 
for the unresolved sources detected in the field. The magnitudes 
and colours were calibrated to the Sloan Digital Sky Survey (SDSS) 
photometric system and corrected
 for interstellar extinction \citep{2011ApJ...737..103S}. 
Objects with integrated colours 
in the ranges 0.5$<(g'-i')_0<$1.4, 0.35$<(g'-r')_0<$0.85 
and 0.0$<(r'-i')_0<$0.55 and magnitudes $g'_0 >22$ were 
considered as GC candidates \citep{2011MNRAS.416..155F}. Three UCD candidates
were selected using the same colour ranges, but with $g'_0 < 22$,  
which translates into $M_V<-11$ mag at the NGC\,5044 distance \citep{2006AJ....131.2442M}.

\subsection{Radial velocity}
\label{rv}

A preliminary estimation of the radial velocity ($V_{\rm r}$) of the three 
UCD candidates was obtained from the spectroscopic data using the task 
\textsf{FXCOR} from IRAF. As spectral templates, we used the stellar
population synthesis models MILES \citep{2010MNRAS.404.1639V} 
considering two total metallicities ($[Z/H]=-1.31$ and $[Z/H]=0.00$ dex), 
a bimodal IMF with a slope of 1.3, and ages of 12 Gyr. For two of these 
objects we have obtained $V_{\rm r}<400$ km s$^{-1}$, meaning that they 
are Milky Way (MW) stars. For the third object 
(hereafter, NGC\,5044-UCD1) we have obtained $V_{\rm r}\sim 2900$ km 
s$^{-1}$, similar to the reported systemic velocity of 
the galaxy NGC\,5044 (2782 km  s$^{-1}$), thereby confirming 
that NGC\,5044-UCD1 is a member of this system.

\begin{figure}
\centering
\includegraphics[scale=0.23]{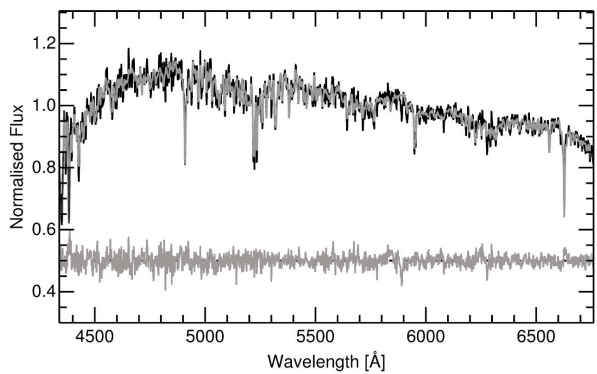}
\caption{Integrated spectrum of NGC\,5044-UCD1 (black) and one of the 
kinematics fits obtained by pPXF (grey). We also show the residuals of the fit 
offset by 0.5, for clarity.}
\label{spectra}
\end{figure}

Subsequently, we used the penalized pixel-fitting 
code pPXF \citep{2004PASP..116..138C} in order to mitigate the mismatch
template effect and, therefore, to better
estimate the $V_{\rm r}$ of the UCD. We used the SSP MILES 
library models from \citet{2010MNRAS.404.1639V}, which span a range 
in age of 0.03-14 Gyr and $-2.27<[Z/H]<0.4$ dex.
We 
obtained a radial velocity of $V_{\rm r}\sim2945$ 
km s$^{-1}$. This value and the error quoted in Table\,\ref{parameters} 
were obtained as the median from 500 Monte Carlo 
realizations and the $1\sigma$ estimation, respectively. 
Figure\,\ref{spectra} shows the integrated spectrum of the
UCD together with an example of the fits obtained by pPXF. 

\subsection{Age and metalicity from Lick indices and SSP models}
\label{Lick}

\begin{figure}
\center
\includegraphics[scale=0.9]{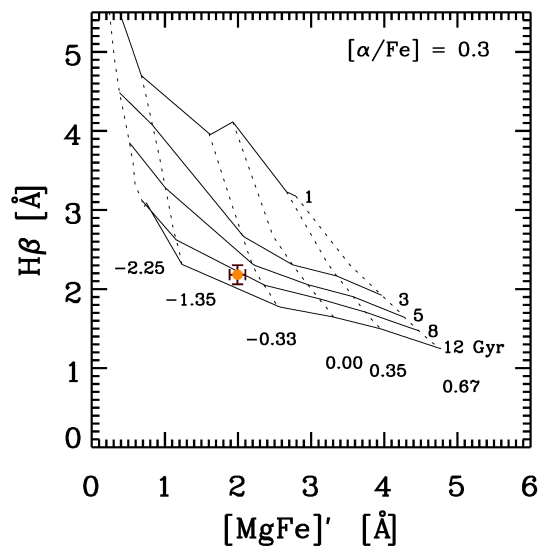}
\caption{Diagnostic diagram $H_\beta$ vs $[Mg Fe]$. 
We also overplot the models of \citet{2003MNRAS.339..897T,2004MNRAS.351L..19T} 
with $[\alpha/Fe]=0.3$. NGC\,5044-UCD1 (filled circle with 
error bars) is an ancient object with subsolar metallicity.}
\label{diagnostic}
\end{figure}

We have obtained the Lick indices from the GMOS spectra following \citet{2008MNRAS.385...40N,2015MNRAS.451.3615N} (Table\,\ref{parameters}). Uncertainties for the indices were obtained from 500 Monte Carlo simulations based on a combination of errors introduced by photon noise (both target and sky) and errors in the measured redshift of the spectra (essentially negligible here). The age, metallicity, and $\alpha$-elements ratio 
of NGC\,5044-UCD1 were obtained through the $\chi^2$ minimization method of 
\citet{2002MNRAS.333..517P} and \citet{2004MNRAS.355.1327P}. 
We have used a smooth grid in age, [$\alpha$/Fe], and [Z/H] by 
interpolating the single stellar 
population (SSP) models of \citet{2003MNRAS.339..897T,2004MNRAS.351L..19T}. 
As in similar works \citep[e.g. ][]{2006MNRAS.367..815N}, the indices $H_{\delta A}$, 
$H_{\delta F}$, $H\gamma A$ 
G4300, Fe4383, $H\beta$, Fe5015, Mgb, Fe5270, Fe5335, and Fe5406 gave the
most reliable results. The analysis provides an age of 
$11.7^{+1.4}_{-1.2}$ Gyr, a total metallicity of $[Z/H]=-0.79\pm0.04$ dex, 
and an $\alpha$-elements ratio of $[\alpha/Fe]=0.30\pm0.06$.
 From these values and using the M/L ratios given 
by \citet{2005MNRAS.362..799M} for a Kroupa IMF ($\rm M/L=2.9\pm0.3$), we infer 
a total stellar mass of $2.8(\pm0.3)\times10^7~M_\odot$.
In  Figure\,\ref{diagnostic} we show the diagnostic diagram $H_\beta$ vs 
$[Mg Fe]$, where $[MgFe]' = [Mgb\times (0.72 \times Fe5270 + 0.28 \times Fe5335)]^{0.5}$, 
overplotting the models of \citet{2003MNRAS.339..897T,2004MNRAS.351L..19T} 
with $[\alpha/Fe]=0.3$, according to the value previously estimated.

\subsection{Star formation history}

Through the full spectral fitting technique implemented within the pPXF code,
we studied the star formation history (SFH) of NGC\,5044-UCD1 and obtained
new estimates of the age and metallicity of this object 
in a way that is 
independent of that presented in Section\,\ref{Lick}.  
We applied the regularization method to obtain a smooth solution 
of the best fit linear combination of SSP models to the UCD spectrum 
\citep{2015MNRAS.448.3484M}. We used the same library of models
presented in Section\,\ref{rv}, adopting an $\alpha$-element 
ratio of $[\alpha/Fe]=0.4$, which is close to the value obtained in the 
Lick index analysis. 

\begin{figure}
\center
\includegraphics[scale=1.1]{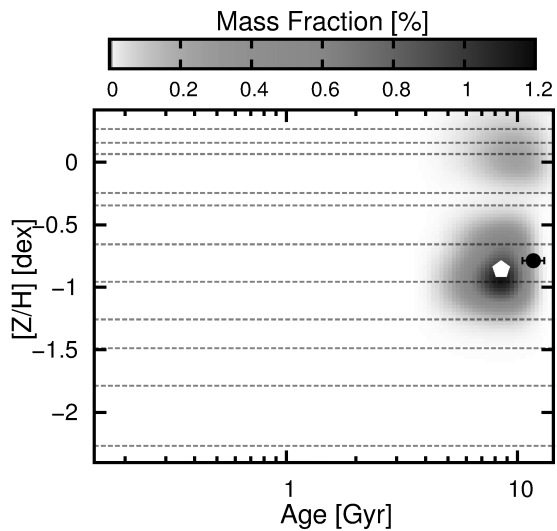}
\caption{Star formation history of NGC\,5044-UCD1 obtained using pPXF. The 
different grey-scale regions indicate
the weight of each SSP model, which is equivalent to the zero-age mass
distribution of the SFH. The dashed lines indicate the $[Z/H]$ of each
available model. The black filled circle depicts the 
luminosity-weighted SSP age and metallicity of this object as measured 
using absorption line strength indices (Section 3.3). The white pentagon 
shows the 
location of the mass-weighted age and metallicity derived from the 
full spectral fitting technique, and considering only models with 
$[Z/H]<-0.25$ dex.}
\label{SFH}
\end{figure}

We performed several pPXF runs considering different spectral ranges of the 
UCD spectrum. In all the cases, we achieved a smooth solution that included a main 
contribution of models around a mean age of 8.52 Gyr and a mean metalicity of 
$[Z/H]= -0.86$ dex (Figure\,\ref{SFH}). In some of the tests, which include
the regions of the highest S/N ($\lambda>4500 \AA$), we also  
obtained a secondary contribution of models with higher metallicities 
($[Z/H]>-0.25$ dex). In Figure\,\ref{SFH} we show one of the tests with  
both contributions. To check this result, we built a synthetic spectrum 
combining two SSP models with the ages and metallicities obtained from our 
analysis for the two components. We added to this spectrum a noise level
similar to that found in the observed spectrum, and we ran pPFX.
We recovered the original two components, but a third
spurious subpopulation was also obtained. To confirm  
this result, we would need a spectrum displaying a higher S/N in the
whole spectral range.

\section{Discussion and conclusions}
\label{Conclusions}

From the photometric and spectroscopic analysis of a deep 
field taken with Gemini+GMOS, we have been able to detect and confirm 
the first UCD in the NGC\,5044 group. Its radial velocity and angular
proximity (2.83 arcmin) indicate that this object is associated with the 
massive early-type galaxy NGC\,5044. At the distance adopted in this 
work for this galaxy, this UCD appears to be an unresolved object, and
it has an absolute magnitude of $M_V=-12.5$ mag. 
This luminosity is well above the usual upper cut for 
``classical'' globular clusters (Faifer et al 2011).

We studied its stellar population properties using two methods: the 
analysis of Lick/IDS indices, and the application of the full spectral 
fitting technique. We find a discrepancy in the age determined by
the two methods. From the comparison of the
Lick/IDS values with SSP models, we have obtained an age of 
$11.7^{+1.4}_{-1.2}$ Gyr, which is significantly older than the value
of 8.5 Gyr obtained with the full spectral fitting technique, although the
uncertainties on the full spectral fitting age are difficult to quantify. Similar 
discrepancies have been reported for UCDs before (e.g. \citealt{2012MNRAS.425..325F}). 
However, in contrast to most of the objects presented by 
Francis and collaborators,
our metallicity values are consistent regardless of the method used
($[Z/H]=-0.79\pm0.04$ dex and $[Z/H]= -0.86$ dex). It is notable
 that the metallicity of NGC\,5044-UCD1 is within the range
displayed by the UCDs detected in Virgo and Fornax, but considerably 
lower than that of the confirmed stripped nuclei presented by 
\citet{2016MNRAS.456..617J}. 

In common with most globular clusters 
(\citealt[][]{2005A&A...439..997P,2008MNRAS.385...40N}), NGC\,5044-UCD1 presents a 
supersolar $\alpha$-element abundance of $[\alpha/Fe]=0.30$, suggesting 
rapid star formation. The SFH analysis suggests the presence of a secondary
metal-rich stellar population, of an age indistinguishable from the main population. 
If confirmed by higher S/N data, this result could be associated with the 
``blue tilt'' phenomena of massive GCs, whereby the most massive low-metallicity 
GCs are found to be progressively redder. This has been interpreted as being 
due to self-enrichment, where the most massive GCs are able to form additional 
generations of stars enriched in metals from the ejecta of the initial population 
(\citealt{2008AJ....136.1828S}).

According to \citet{2012MNRAS.425..325F}, the UCDs of Virgo 
and Fornax do not display properties similar to those of the 
nuclei of present-day nucleated dwarf galaxies. This picture seems to be
in agreement with the location that our UCD would have in Figure\,13 of 
\citet{2011MNRAS.414..739N}. In that figure, the absolute visual magnitude of our object 
($M_V=-12.5$ mag) 
and the total number of GCs of NGC\,5044 ($\sim5500 \pm 500$, Faifer et al. in 
preparation) would agree with the general trend defined by the UCDs that are more 
likely the brightest tail of the GC population. In this sense, from a
Monte Carlo experiment using the luminosity function of the NGC\,5044 GC
system, we have obtained that in 30\% of the realizations
we have $1-3$ GCs brighter than $M_V=-12.5$ mag.
Therefore, all the lines of evidence would point toward NGC\,5044-UCD1 
being an unusually massive GC of the NGC\,5044 system.

\begin{acknowledgements}
Based on observations obtained at the Gemini Observatory 
(Programs GS-2009A-Q-46 and GS-2010A-Q-56), 
which is operated by the Association of Universities for Research in 
Astronomy, Inc., under a cooperative agreement with the NSF on behalf of the 
Gemini partnership: the National Science Foundation 
(United States), the National Research Council (Canada), CONICYT (Chile), the 
Australian Research Council (Australia), Minist\'erio da Ci\^encia, Tecnologia 
e Inova\c c\~ao (Brazil) and Ministerio de Ciencia, Tecnolog\'ia e 
Innovaci\'on Productiva (Argentina).\\

This research has made use of the NASA/IPAC Extragalactic Database (NED) which is 
operated by the Jet Propulsion Laboratory, California Institute of Technology, 
under contract with the National Aeronautics and Space Administration.\\

This work was funded with grants from Consejo Nacional de Investigaciones 
Cient\'{\i}ficas y T\'ecnicas and Universidad Nacional de La Plata
(Argentina).

\end{acknowledgements}

%
%
\bibliographystyle{aa}
\bibliography{Faifer_et_al}

\end{document}